\documentclass[12pt]{iopart}
\usepackage{citesort}
\usepackage{graphicx}
\usepackage{iopams}

\makeatletter
\@namedef{ver@amsmath.sty}{}
\makeatother

\usepackage[version=4]{mhchem}
\usepackage{chemformula}
\usepackage{CJK}
\begin{document}

\title[]{In silico investigation of Ba-based ternary chalcogenides for photovoltaic applications}

\author{Ramya Kormath Madam Raghupathy, Hossein Mirhosseini, and Thomas D. K{\"u}hne}
\address{Dynamics of Condensed Matter and Center for Sustainable Systems Design, Chair of Theoretical Chemistry, University of Paderborn, Paderborn, German}
\ead{h.mirhosseini@hzdr.de}

\begin{abstract}
In solar cells, the absorbers are the key components for capturing solar energy and converting photons into electron-hole pairs. The search for high-performance absorbers with advantageous characteristics is an ongoing task for researchers. In this work, we investigated promising and environmentally benign Ba-based ternary chalcogenides for photovoltaic applications. The total number of Ba-based ternary chalcogenides in the Materials Project database was found to be 279. Materials screening based on bandgap size and stability reduced the number of compounds to 19. The performance of an absorber depends on the charge carrier lifetime, which is controlled by non-radiative processes involving defects. Hence, we investigated the intrinsic defects and p-type dopability of the compounds. We identified two Ba-based compounds, namely \ch{BaCu2Se2} and \ch{ZrBaSe3}, as promising absorbers for single-junction and tandem cells and investigated them in detail.
 \end{abstract}


\section{Introduction}
There is currently a growing demand for energy resources with low environmental impact, which presents a real challenge. The common goal is to find sustainable energy alternatives with near-zero carbon footprint~\cite{Gielen2019}. In this regard, the abundant availability of solar energy makes photovoltaic (PV) technology particularly attractive~\cite{Green2016}. The key component in any solar cell is the absorber material. Currently, the most commonly used absorber is multicrystalline silicon owing to the relatively low cost and high efficiency of Si solar panels~\cite{ELLABBAN2014748,PARIDA20111625,Huang_2021}. In addition, chalcogenide-based absorbers, such as Cu(In,Ga)Se$_2$ (CIGSe) and Cu$_2$ZnSn(S,Se)$_4$, as well as emerging absorbers like perovskites are currently being studied for application in thin-film solar panels~\cite{ELLABBAN2014748,PARIDA20111625,Huang_2021,Jackson2015,Tian2020}. Thin-film technologies are of particular interest because they have the potential to substantially reduce the cost of clean energy generation by reducing the amount of material required to fabricate thin-film solar cells~\cite{PARIDA20111625,HUSAIN2018779,POWALLA2017445}. 

Significant progress in PV technology has been made with solar cells based on CIGSe, CdTe, and perovskites, which have demonstrated over 20\% lab-scale power conversion efficiency~\cite{Adrian2013,Jackson2016,C4EE01389A,Polman2016,NREL}. However, the commercialization of cells based on these absorbers is impeded by the scarcity (e.g. In, Ga, and Te) or toxicity (e.g. Pb and Cd) of their constituent elements as well as the long-term stability issue of halide-based perovskites~\cite{Polman2016,WOODHOUSE2013199,C3TA13655H,Shin2017,KALAISELVI2018198}.

For thin-film solar cells, several non-toxic, cheap, and earth-abundant chalcogenides have been proposed over the past few years, such as \ch{SnS}, \ch{GeSe}, Sb$_2$(S,Se)$_3$, \ch{CuSbSe2}, and \ch{Cu2SnS3}. However, the record efficiency of cells based on these materials is below 10\%, far below the necessary efficiency for practical applications~\cite{C9QM00727J,D0TA06937J,MESSINA20092503,GUO2019586,RAMPINO201886,Nakashima_2015,MAVLONOV2020227}.

Recently, a new class of materials, chalcogenide-based perovskites, has been extensively investigated~\cite{Buffiere2019,Abhishek2019,Sun_2015,Osei2019}. Nishigaki and co-workers fabricated chalcogenide-based perovskite polycrystals with an ABS$_{3}$ formula (A = Ba, Sr; B = Zr, Hf) and found that Ba(Zr,Ti)S$_{3}$ alloys have great potential for tandem solar cell applications~\cite{Nishigaki2020}. Based on first-principles calculations, Sun \emph{et al.} determined the electronic structure for a wide variety of transition metal chalcogenides of the form ABX$_{3}$ (A = Ca, Sr, Ba; B = Ti, Zr, Hf; X = S, Se) assuming a set of crystal geometries~\cite{Sun_2015,Sun2019}. Their calculations suggested that the substitution of O with S or Se lowered the bandgap close to the range required for solar cell application. Similarly, Ong and co-workers performed first-principle calculations and ab-initio molecular dynamics simulations to probe the electronic structure and stability of BaZr(S,Se)$_{3}$ and found that \ch{BaZrSe3}, with a small bandgap and high optical absorption coefficients, is likely to be stable at room temperature~\cite{Marc2019}. Although experimental and theoretical work has been done on chalcogenide-based perovskites, their electronic and optical properties under different physical and chemical conditions still require elucidation~\cite{PERERA2016129,Clearfield:a03737,Niu2017,ZITOUNI2020110923,PhysRevMaterials.4.091601,D0TC04516K}.

To meet the requirements of next-generation PV devices, alternative absorbers that consist of earth-abundant, cheap, and non-toxic elements are required. Recently, the standard technique for finding promising materials for targeted applications has been computational materials screening based on first-principle calculations~\cite{Huang_2021,Hautier2021,yu2012,yu2013,fabini2019,Pandey2017,ramya,KormathMadamRaghupathy2018h,Wiebeler_2020}. In this work, we performed first-principle calculations to find Ba-based chalcogenides that show promise as light absorbers. The choice of materials family stemmed from the abundance and lower toxicity of Ba compared to the constituent elements of CIGS, CdTe, and lead-containing perovskites~\cite{WinNT}. In addition, solar cells based on BaCu$_{2}$Sn(S,Se)$_{4}$ absorbers have demonstrated 5\% efficiency~\cite{Xiao2017,Kumar_2020,Luo2020}. To design efficient solar cells, a fundamental understanding of the properties of the new compounds is crucial. Therefore, we studied the electronic and optical properties of promising Ba-based ternary chalcogenides. Moreover, we investigated the diffusion of point defects by employing the climbing-image nudged elastic band (CI-NEB) method \cite{Henkelman2000}
\section{Methodology}
All calculations were performed within the framework of density functional theory using the Vienna Ab-initio Software Package (VASP)~\cite{vasp}. We used projector augmented wave (PAW) pseudopotentials with a plane-wave cutoff of 500 eV~\cite{paw}. The Perdew-Burke-Ernzerhof (PBE) form of the generalized gradient approximation was employed for the exchange-correlation potential~\cite{pbe}. Atomic structures were considered to be optimized when the residual force on each atom was less than 0.01 {eV/$\text{\AA}$}. Brillouin zone integration was performed on a
Monkhorst-Pack mesh centered at the $\Gamma$ point~\cite{Monkhorst_1976}. The density of k-point was kept constant for all systems by specifying the `KSPACING' tag. The electronic structures for the selected compounds were calculated using the HSE06 screened hybrid functional~\cite{heyd_JPC_2003}. For modeling of point defects, we used the PyCDT toolkit \cite{pycdt}, which automatically creates files for the  bulk, dielectric, and defect calculations. Defect calculations were performed in two steps: first, we computed defect formation energies at the PBE level and then we performed HSE06 calculations for the compounds that showed promising p-type dopability. Defect formation energies and corresponding charge transition levels were calculated for large supercells to minimize finite-size effects. For more details see Ref. \cite{Mirhosseini2020}. 
The optical absorption spectrum was computed from hybrid calculations using the independent particle approximation~\cite{Gajdos2006}. 
To model the diffusion of point defects, we employed the CI-NEB method \cite{Henkelman2000}. A linear interpolation technique was applied to create five equidistant images for each NEB calculation. The force constant of the spring, which connects the images, was set at 5.0 eV/${\rm{ \AA^2}}$. 

\section{Results and discussion}
Our previous materials screening search led to the identification of binary and ternary compounds that are promising as p-type transparent conductors~\cite{ramya,KormathMadamRaghupathy2018h,Wiebeler_2020}. Here, we performed a first-principle materials screening to identify Ba-based ternary chalcogenides as suitable p-type absorbers.
The total number of Ba-based ternary chalcogenides available in the Materials Project database (MPDB)~\cite{materials} was found to be 279. After excluding compounds containing toxic elements like Pb and  Cd, less abundant elements like Ga and In, and expensive elements as well as lanthanides and actinides, the number of Ba-based ternary chalcogenides with E$_{hull}$ smaller than 100 meV/atom was found to be 25. In the next step, HSE06 electronic structure calculations were performed for compounds having a PBE bandgap between 0.1 eV and 1.0 eV. The final 19 compounds with bandgaps between 1.0 eV and 2.0 eV (calculated with the hybrid functional) are tabulated in Table~\ref{tbl:list}.
%
%
\begin{table}[ht]
\caption{\label{tbl:list} List of Ba-containing ternary chalcogenides with bandgaps between 1.0 eV and 2.0 eV calculated with the hybrid functional. For each compound, the MPDB ID, symmetry, and HSE06-calculated bandgap (\(\mathrm{E^{HSE}_{g}}\)) are shown.}
\footnotesize
\begin{center}

\begin{tabular}{lllccc}
\hline
\hline
Compound & MPDB ID &  Space group & \(\mathrm{E^{HSE}_{g}}\)  \\
\hline
\hline 
\ch{Ba(PdS$_{2}$)$_{2}$} & mp-28967 & P2$_{1}$/m & 1.87 \\
\ch{Ba$_{2}$Ag$_{8}$S$_{7}$} & mp-1228614 & Pmn2$_{1}$ & 2.00   \\
\ch{BaSnS$_{3}$} & mp-1183370 & Pnma & 1.72   \\
\ch{Ba$_{3}$Zr$_{2}$S$_{7}$} & mp-9179 & I4/mmm & 1.26   \\
\ch{Ba$_{3}$Zr$_{2}$S$_{7}$} & mp-8570 & P4$_{2}$/mnm & 1.26   \\
\ch{Ba$_{3}$Zr$_{2}$S$_{7}$} & mp-554172  &   Cccm & 1.26 \\
\ch{Ba$_{2}$ZrS$_{4}$} & mp-3813  &   I4/mmm & 1.37  \\
\ch{Ba(AgS)$_{2}$} & mp-8579  &   P3m1 & 1.90  \\
\ch{Ba$_{4}$Zr$_{3}$S$_{10}$} & mp-14883  &   I4/mmm & 1.27  \\
\ch{Ba(BiS$_{2}$)$_{2}$} & mp-28057  &  P6$_{3}$/m & 1.58  \\
\ch{BaCu$_{4}$S$_{3}$} & mp-654109 &  Pnma & 1.28  \\
\ch{Ba(CuS)$_{2}$} & mp-4255 &   I4/mmm & 1.80  \\
\ch{BaPdS$_{2}$} & mp-4009  &   Cmcm & 1.61  \\
\ch{BaAg$_{8}$S$_{5}$} & mp-29682  &  P2$_{1}$/m   & 1.41 \\
\ch{Ba(SbSe$_{2}$)$_{2}$} & mp-4727 &   P2$_{1}$/c  & 1.70  \\
\ch{BaBiSe$_{3}$} & mp-27365 & P2$_{1}$2$_{1}$2$_{1}$ & 1.20  \\
\ch{Ba(CuSe)$_{2}$} & mp-4473 & Pnma & 1.70  \\
\ch{Ba(CuSe)$_{2}$} & mp-10437 & I4/mmm & 1.38  \\
\ch{BaZrSe$_{3}$}  & mp-998427 & Pnma & 1.00 \\
\hline 
\hline
\end{tabular}
    
\end{center}
\end{table}

A suitable bandgap and good carrier effective mass are the primary requirements for a high-performance absorber. However, it has been shown that semiconductors with suitable bandgaps and carrier effective masses might exhibit low efficiency in optoelectronic applications due to deep defect states~\cite{Ming2016}. Recently, Hautier and co-workers performed high-throughput screening calculations to identify defect-tolerant Cu-based solar absorbers~\cite{Hautier2021}. Their results suggested that exploring the nature of defects in absorber materials is critical.

Among the 19 Ba-based chalcogenides tabulated in Table~\ref{tbl:list}, only a few compounds were found to possess acceptable characteristics for p-type dopability. It was observed that vacancy and interstitial defects can be hole killer defects in the majority of the compounds. In addition, efforts to find suitable extrinsic dopants to further enhance p-type dopability were not successful for some of the compounds. For example, for BaPdS$_{2}$ and BaPd$_{2}$S$_{4}$, most of the defects were found to be neutral, but the enhancement of hole conductivity by dopant incorporation was found to be unfeasible for these compounds.

For BaCu$_{4}$S$_{3}$ (Pnma), the vacancy defects Vac$_{\rm Cu}$ and Vac$_{\rm Ba}$ were found to have low formation energies compared to other intrinsic defects under anion-rich conditions. These defects can introduce holes in the valence band and enhance hole conductivity. In agreement with our results, it has been reported that small amounts of Vac$_{\rm Ba}$ and Vac$_{\rm Cu}$ defects are the origin of high p-type conductivity in BaCu$_{4}$S$_{3}$~\cite{Han2017}. To further enhance hole concentration, we attempted to dope this compound with Cs and Zn. However, the formation energies of the extrinsic defects were found to be higher than those of the intrinsic defects, so the formation of these extrinsic defects is energetically unfavorable. Similarly, the formation of Vac$_{\rm Ag}$ in BaAg$_{8}$S$_{5}$ could enhance hole concentration, but extrinsic defects were found to have higher formation energies compared to intrinsic defects. These two materials are intrinsically p-type conductors, however, we could not find a dopant for them. 

Our screening based on defect chemistry led to the identification of 4 promising candidates with good p-type dopability, namely ${\beta}$-BaCu$_{2}$S$_{2}$ (I4/mmm), $\alpha$-BaCu$_{2}$Se$_{2}$ (Pnma), ${\beta}$-BaCu$_{2}$Se$_{2}$ (I4/mmm), and \ch{BaZrSe3} (Pnma). Among these compounds, ${\beta}$-BaCu$_{2}$S$_{2}$ and ${\beta}$-BaCu$_{2}$Se$_{2}$ are high-temperature phases with bandgaps of 1.75 eV and 1.38 eV, respectively~\cite{Sangmoon2002,Krishnapriyan_2014}. These compounds have been found to be promising thermoelectric materials with a high Seebeck coefficient and high p-type conductivity~\cite{Han2017}. Our results also indicated the promising p-type dopability of these two compounds, as the Vac$_{\rm Ba}$ and Vac$_{\rm Cu}$ defects were found to have low formation energies compared to other defects. Furthermore, we found that alkali-induced defects are favorable for p-type conductivity.

\begin{figure*}[ht]
\centering
\includegraphics[scale=0.45]{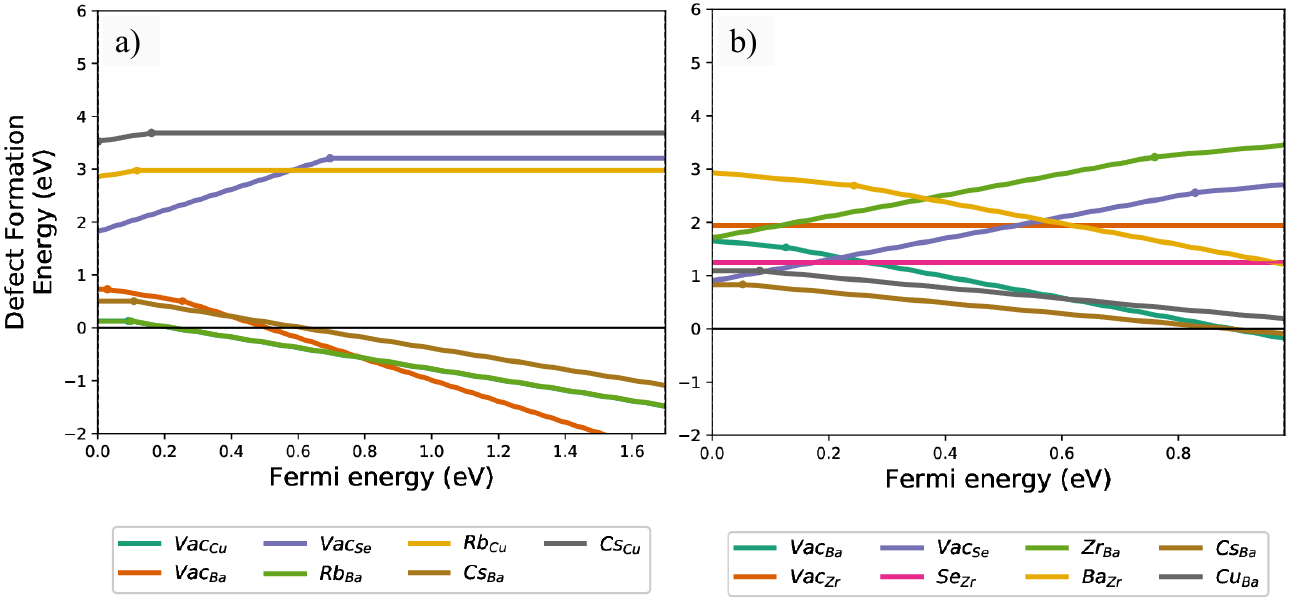}
\caption{The defect chemistry for a) $\alpha$-\ch{BaCu2Se2} and b) \ch{BaZrSe3} calculated with the hybrid functional.}
\label{fig:defects}
\end{figure*}

The low temperature $\alpha$-BaCu$_{2}$Se$_{2}$, with a high hole carrier mobility (10-15 cm$^{2}$V$^{-1}$s$^{-1}$), has been experimentally characterized for thermoelectric applications~\cite{Li2015}. The hybridization of Se-4p and Cu-3d states at the valence band maximum leads to a small hole effective mass, enhancing electrical conductivity. The carrier concentration of undoped $\alpha$-\ch{BaCu2Se2} was found to be 1.7 x 10$^{18}$ cm$^{-3}$~\cite{Li2015}. In addition, it has been shown that Na doping can increase the electrical conductivity of $\alpha$-\ch{BaCu2Se2} by two orders of magnitude~\cite{Li2015}. From the defect physics (see Figure~\ref{fig:defects}a), we found that Vac$_{\rm Cu}$ and Vac$_{\rm Ba}$ are shallow acceptor defects and have lower formation energies compared to Vac$_{\rm Se}$. It suggests that these defects can enhance hole concentration, in agreement with  experimental observations. The antisite as well as interstitial defects have high formation energies and are difficult to form. The substitutional defects Rb$_{\rm Ba}$ and Cs$_{\rm Ba}$ were found to increase hole concentration in $\alpha$-\ch{BaCu2Se2}, as shown in Figure~\ref{fig:defects}a.  

Defect formation energies for \ch{BaZrSe3} under anion-rich conditions are shown in Figure~\ref{fig:defects}b. Out of all the intrinsic defects, Vac$_{\rm Se}$ (donor) and Se$_{\rm Zr}$ (neutral) have the lowest formation energies when the Fermi level is close to the valence band maximum. Regarding the dopants, we observed that Cs$_{\rm Ba}$ defects have lower formation energy compared to Zr$_{\rm Ba}$ and Vac$_{\rm Ba}$ intrinsic defects, whereas Cu$_{\rm Ba}$ is more stable than Vac$_{\rm Ba}$. The strong localization of the upper edge of the valence band stems mainly from the relatively localized Se 4p orbital, leading to a less disperse valence band. 

\begin{figure*}[ht]
\centering
\includegraphics[scale=0.45]{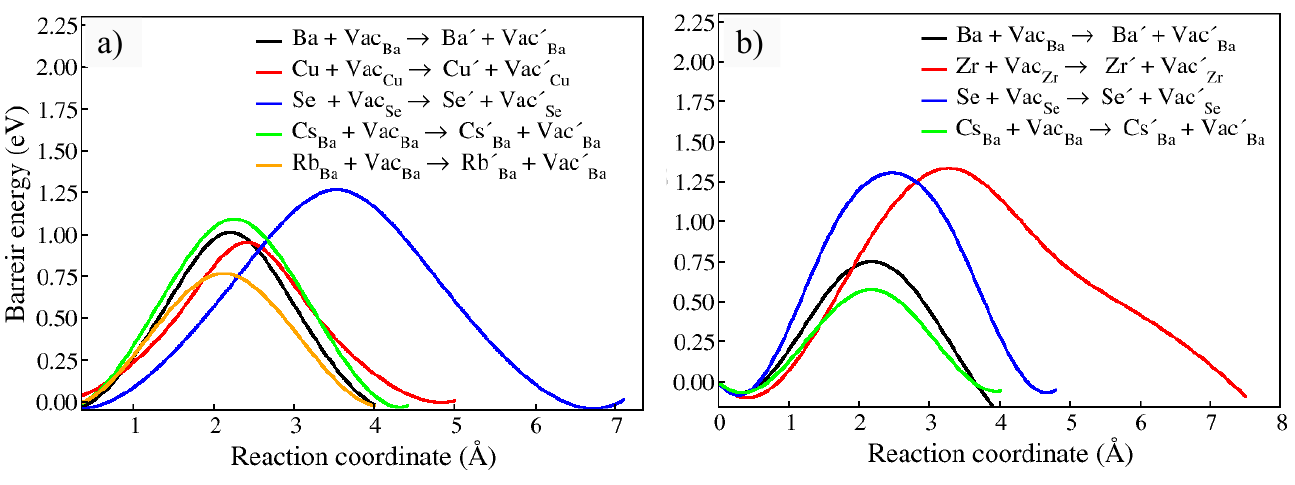}
\caption{Migration barriers for the vacancy-mediated diffusion of intrinsic and extrinsic defects in a) $\alpha$-\ch{BaCu2Se2} and b) \ch{BaZrSe3}.}
\label{fig:neb}
\end{figure*}

As shown, vacancy defects have low formation energies in BaCu$_{2}$Se$_{2}$ and \ch{BaZrSe3}. Therefore, we considered vacancy-mediated diffusion mechanisms for mass transport in these compounds. Our results showed that the migration barriers for Vac$_{\rm Cu}$ and Vac$_{\rm Ba}$ in BaCu$_{2}$Se$_{2}$ are smaller than the migration barrier of Vac$_{\rm Se}$. When alkali atoms are introduced into the system, they can occupy different sites in the BaCu$_{2}$Se$_{2}$ lattice, depending on their atomic radii. The formation energies indicated that alkali atoms like Rb and Cs can incorporate into BaCu$_{2}$Se$_{2}$ by forming Ba-substitutional defects. The migration barrier for Rb$_{\rm Ba}$ was comparatively smaller than Vac$_{\rm Ba}$, whereas the migration barrier for Cs$_{\rm Ba}$ was similar to that of Vac$_{\rm Ba}$. The migration barriers for Vac$_{\rm Zr}$ and Vac$_{\rm Se}$ in \ch{BaZrSe3} were relatively large (see Figure~\ref{fig:neb}b) compared to the migration barrier for Vac$_{\rm Ba}$. The formation energies indicate that Cs can incorporate into the \ch{BaZrSe3} lattice through the formation of Cs$_{\rm Ba}$ substitutional defects. The migration barrier for Cs$_{\rm Ba}$ was smaller than those for intrinsic defects.

\begin{figure}[ht]
\begin{center}
\includegraphics[scale=0.5]{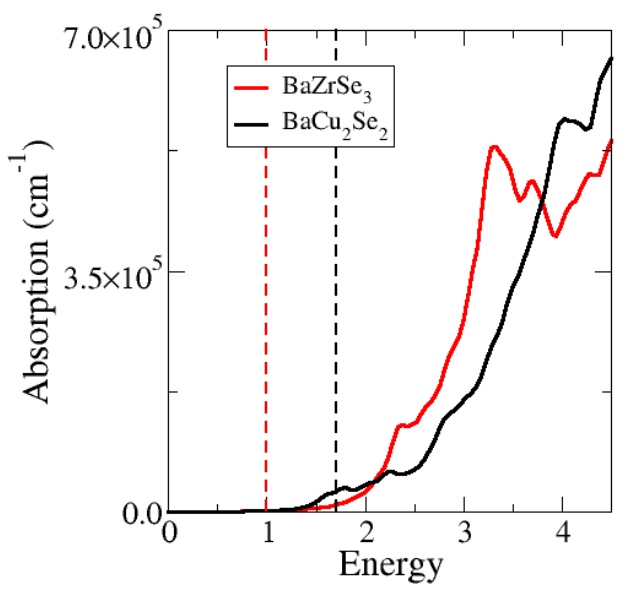}
  \caption{Optical absorption spectra for $\alpha$-\ch{BaCu2Se2} and \ch{BaZrSe3}. Dotted lines indicate corresponding bandgaps.}
\label{fig:opticalspectra}   
\end{center}
\end{figure}

An important characteristic to be a high-performance absorber is an optimal direct bandgap~\cite{C7EE00826K}. Optical absorption is another important property for solar cell absorbers, because a direct bandgap does not guarantee the absorption in the visible spectrum needed~\cite{ju2017}. For example, typical silicon absorbers have to be hundreds of micrometers in thickness due to the indirect bandgap and low absorption coefficient of Si~\cite{ju2017}. In contrast, the thickness of CIGSe absorbers can be about 100 times less than that of Si absorbers owing to the direct bandgap and the high absorption coefficient of CIGSe~\cite{LI2019704}. The promising compounds identified here are predicted to possess direct bandgaps at $\Gamma$ point; a characteristic that makes them more suitable for thin-film solar cells. Additionally, both compounds exhibit strong absorptions in the visible spectrum (see Figure~\ref{fig:opticalspectra}) which are comparable with those of CuInSe$_{2}$ and MAPbI$_{3}$~\cite{HORIG197867,ju2017}.
\section{Conclusion}
 We performed material screening search based on first-principle calculations to identify promising Ba-based ternary chalcogenides available in the MPDB as p-type light absorbers. Our screening descriptors are based on the bandgap size, optical absorption spectra, and defect chemistry. In the search of absorbers for use in single-junction and tandem cells, we found two promising compounds that are earth-abundant, stable, and less toxic than existing absorbers. Our results indicated that \ch{BaZrSe3} has a suitable bandgap and good p-type dopability and thus might be used in  single-junction cells or as a bottom cell absorber in tandem cells. In addition, $\alpha$-\ch{BaCu2Se2}, which has a bandgap of 1.7 eV and good p-type dopability, was found to be an ideal candidate for the top cell absorber in tandem cells. We anticipate that our results will inspire experimental research groups to synthesize Ba-based compounds as alternative p-type absorbers.

\section*{Acknowledgments}
The authors would like to acknowledge financial support from the German 
\emph{Bundesministerium f\"ur Wirtschaft und Energie (BMWi)} for the speedCIGS project (0324095C). Ramya Kormath Madam Raghupathy acknowledges the University of Paderborn for Postdoc scholarship.  %
The authors would like to acknowledge the Paderborn Center for Parallel Computing (PC$^{2}$) supercomputing time on Noctua. The authors gratefully acknowledge the Gauss Centre for Supercomputing e.V. (www.gauss-centre.eu) for funding this project by providing computing time through the John von Neumann Institute for Computing (NIC) on the GCS Supercomputer JUWELS at Jülich Supercomputing Centre (JSC).

\section*{References}
\bibliography{Ba-based-absorber}
\end{document}